# Arianna: towards a new paradigm for assistive technology at home

Luca Buoncompagni[1], Barbara Bruno[2], Antonella Giuni[3], Fulvio Mastrogiovanni[4] and Renato Zaccaria[5]

*Abstract.* Providing elderly and people with special needs to retain their independence as long as possible is one of the biggest challenges of the society of tomorrow. Teseo, a startup company spinoff from University of Genoa, aims at accelerating the transition towards a sustainable healthcare system. Teseo's first concept and product, Arianna, allows for the automated recognition of activities of daily living at home, and acts as a wellbeing and healthcare personalised assistant. This abstract outlines the main concepts underlying its features and capabilities.

## 1 Motivation

The mission of Teseo is allowing elderly and people with mild cognitive impairments to retain their independence as long as possible, while living in their home.

The first step for Teseo is to develop and market an easy to use and cheap modular device kit, called Arianna, to assist elderly people and those suffering from cognitive or physical impairments.

The *home care* model is spreading in all industrialized countries. It addresses two needs: (i) supporting elderly and people with special needs, both in a post-hospitalization phase and when it is necessary to have a mid- or long-term care service; (ii) helping people who do not have access to home care services. In both cases, it is necessary to ensure that the Activities of Daily Living (ADL) are correctly and regularly performed. ADLs are daily activities related to motion, rest, nutrition, and personal hygiene, which measure a person's wellbeing, determine their life quality and independence level.

---

[1] Luca Buoncompagni is with the Department of Informatics, Bioengineering, Robotics and Systems Engineering, University of Genoa. Email: luca.buoncompagni@edu.unige.it

[2] Barbara Bruno is with the Department of Informatics, Bioengineering, Robotics and Systems Engineering, University of Genoa, and Teseo srl. Email: barbara.bruno@unige.it

[3] Antonella Giuni is with Teseo srl. Email: antonella.giuni@teseotech.com

[4] Fulvio Mastrogiovanni is with the Department of Informatics, Bioengineering, Robotics and Systems Engineering, University of Genoa, and Teseo srl. Email: fulvio.mastrogiovanni@unige.it

[5] Renato Zaccaria is with the Department of Informatics, Bioengineering, Robotics and Systems Engineering, University of Genoa. Email: renato.zaccaria@unige.it

Teseo's work is based on new approaches to the management of chronic diseases such as cognitive decline [1, 2]. The idea is to adopt personalized and multi-therapeutic approaches. ADLs are particularly relevant to such treatments. These studies show that adopting a proper lifestyle is an essential step in the management and in some cases the regression of disabling chronic diseases.

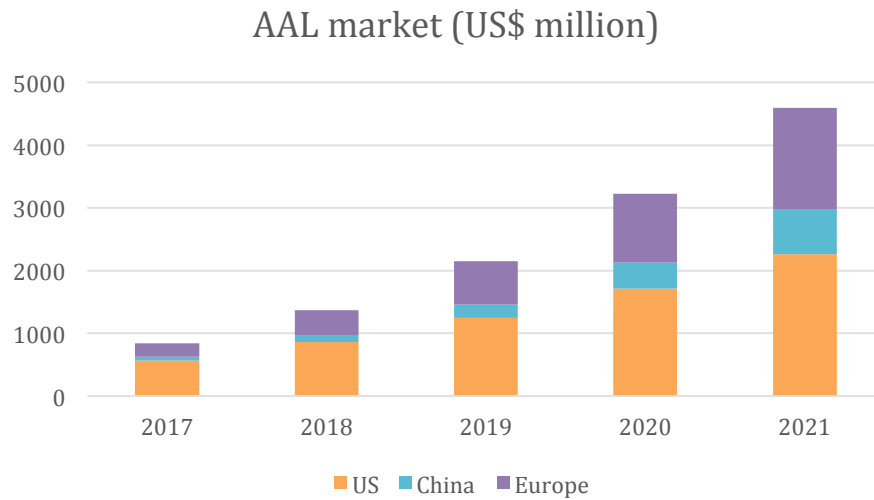

Figure 1: AAL market estimate in 2017-2021

The key market is Ambient Assisted Living (AAL), valued today approximately $1 billion, with a 55.6% CAGR over the 2017-2021 period. Arianna is a set of *nearables* to be installed in an apartment, plus a wearable to be worn or fit in garments. While there are products for specific activities (e.g., nutrition, falls, fitness) or indoor localisation, there is no evidence of *smart* commercial products to detect an important set of ADLs as Arianna can do. After an initial setup phase, Arianna can:

- *localise* people in their home;
- *identify* their significant gestures and correlate them to position and time of day;
- *determine* people activities related to ADLs;
- *interact* with assisted people through dialogue;
- *remind* people, by voice interaction, to perform typical ADLs, if not detected or performed too rarely, acting as a personal assistant;
- *check* their posture, to allow for a quick intervention in case of falls or fainting;
- *learn* their habits and identify anomalous situations;
- *automatically notify* anomalous situations to a relatives, friends, or medical staff.

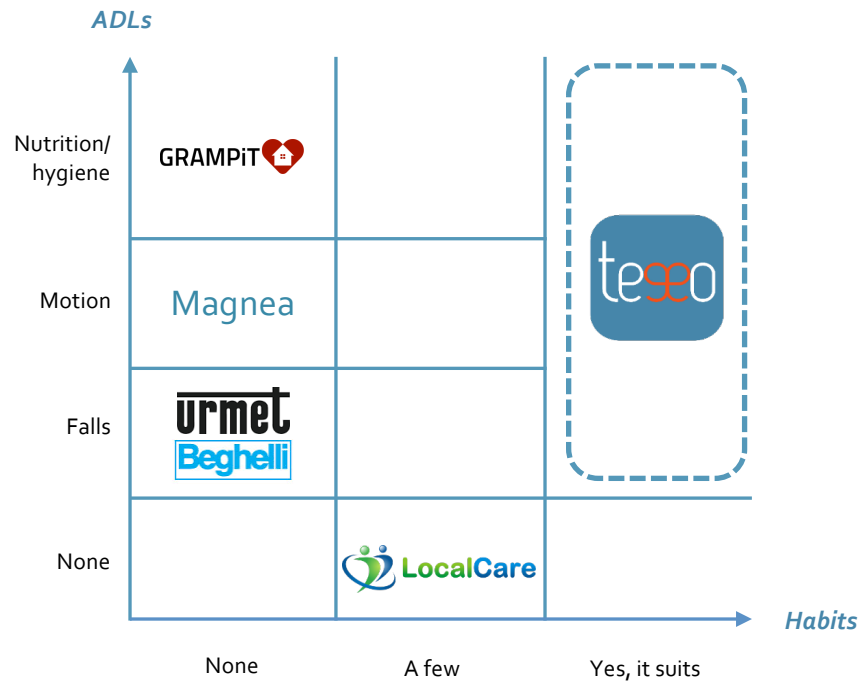

**Figure 2:** Competitor analysis

By means of assistance mechanisms and voice interaction, Arianna encourages assisted people to acquire healthier lifestyles and determines their evolution over time.

## 2 Market analyses

### 2.1 Demand analysis

According to Eurostat, in 2013, the 65+ EU population represented about 18% of the total. According to estimates for 2025, an increase is expected to 25-30%. Comparable statistics are available for many G20 countries. While in the past six decades the world's elderly population increased at the rate of 8-10%, by 2050 the rate is expected to uproar to 22%. According to a study commissioned to the US Census Bureau by the National Institute on Aging (NIA), by 2050 the elderly population will reach 1.6 billion, or nearly 17% of the world's population [3]. Even though in the most developed countries ageing indexes [4] are already above 100 (e.g. Japan 204.9, Italy 161.1, Germany 159.9, as indicated in [5]), the vast majority of elderly are in less developed countries, which also have a faster population ageing. Such demographic trend will raise healthcare costs, in particular of those related to chronic diseases, and will put at risk the sustainability of the welfare state. Home care can be a sustainable solution, both from the economical and human point of view, and could help to keep under control healthcare and welfare budgets in the mid-long term.

*2.2 Competitor analysis*

In recent years, several wearables, such as watches and fitness devices and smart clothing items have been presented or marketed. All products are integrated with smartphones and tablets, generally used for data processing. These devices, however, target only fitness or motion detection.

Even competitors focused on AAL offer partial solutions (Figure 2). Most of them identify only falls, e.g., Italian companies such as Beghelli and Urmet, and American companies, mostly startups. Magnea (Sweden) identifies only motion. GRAMPiT (Italy) uses PIR and magnetic sensors to interpret human actions, but with their approach many false positives are detected. LocalCare (Italy) has developed a solution based on PIR for people with Alzheimer's disease, which can determine a few of their habits.

## 3 Arianna capabilities and user experience

Arianna has a number of key functional capabilities in AAL scenarios:

- Monitoring people locations in the environment at the *topological* level, e.g., *in the kitchen*, *in the bedroom*, *in the bathroom*.
- Monitoring postures and transitions between postures, e.g., *standing*, *seated*, *sitting down*, and *falls*.
- Monitoring a selected number of important ADLs based on gesture detection, e.g., *walking*, *drinking*, *using fork and knife*, *teeth brushing*.
- Speech-based dialogues between Arianna and assisted people to actively obtain information about their state, e.g., *determining if an unusual motion patterns corresponds to a sudden illness*.
- Motion analysis and detection of special sequences of activities, e.g., *sitting for a long time*, *staying still for a long time*, *cooking*, *going often to the bathroom*.

Apart from falls, detected on the wearable, reasoning is performed by the cloud-based Arianna AI engine (Figure 3). Teseo is already working on future Arianna capabilities:

- Activity analysis to fit models of cognitive disability, e.g., to check for positive correlation with such chronic diseases as the Alzheimer's disease.
- Adaptation of monitoring schemes over time, e.g., to evaluate the degradation or recovery of physical or cognitive functions.
- Motivational speech-based interaction, to encourage people to carry out meaningful activities and the ADLs.

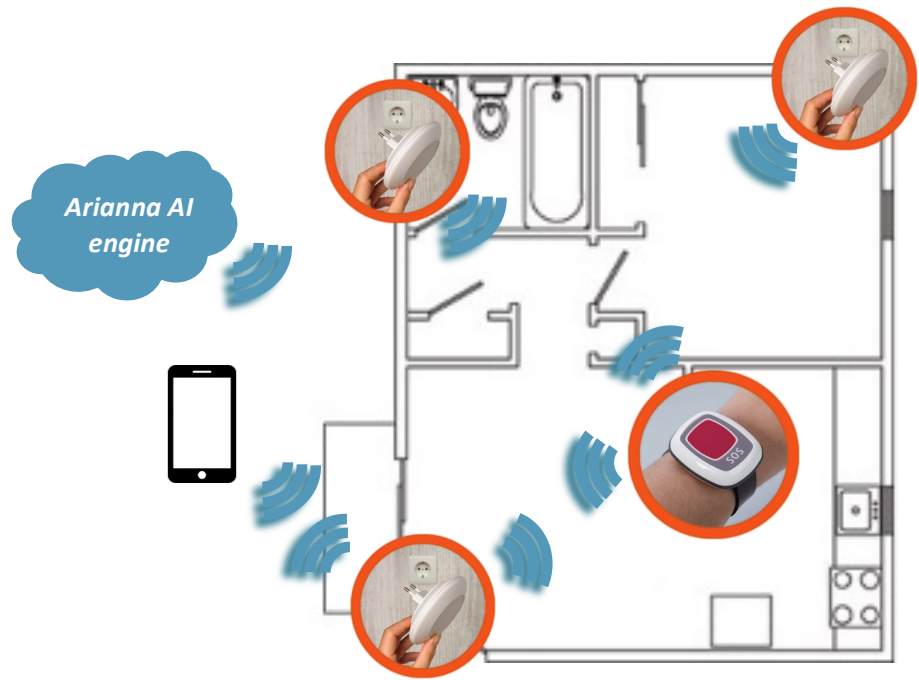

**Figure 3: Arianna's architecture**

Arianna adopts two interaction strategies.

The first is based on a silent and sensible monitoring of what a person does, and Arianna acts as a personalised well-being coach. On the one hand, through the communication between the wearable and nearables, Arianna determines where people are and tracks their movements over time. Using spatial reasoning capabilities, Arianna determines whether people behave in an *anomalous* way: the assisted person never went to the kitchen during the day, or was detected there for too short a period, and therefore may not have had lunch; or the person went to the bathroom much more often than usual, and therefore may be sick. Arianna can interact with people through dialogues, to check on their conditions and act if necessary. On the other hand, the wearable allows Arianna to evaluate whether assisted people perform activities correlated to ADLs, e.g., drinking, teeth brushing, walking regularly. If ADLs are not detected or detected too rarely, Arianna can motivate assisted people to carry out them, e.g., reminding to drink more often and checking whether the action is executed.

The wearable also allows for the detection of accidental falls. The cases in which an assisted person sits on a chair or falls on the floor can give rise to an interpretation ambiguity due to the similarity between inertial data patterns. The ambiguity can be solved by the person's behaviour after the event, for instance if normal activity patterns are detected afterwards. To resolve ambiguities and avoid false alarms, Arianna starts a speech-based dialogue interaction with people, asking simple questions they can answer, typically affirmatively or negatively. Depending on the provided answers or their absence, a request for help to the *outside world* may be invoked, typically through various channels (phone call, SMS, other forms of messaging).

The second interaction strategy assumes the assisted person to voluntarily request assistance by performing a given gesture, for example by issuing a few taps on the wearable surface. The person can start speech-based interaction with Arianna using specific utterances, e.g., "Hello Arianna!"

In order to avoid lack of monitoring because of people not wearing the device, e.g., if they leave the wearable abandoned in some place, Arianna can understand whether it is worn by analysing the associated inertial data pattern, and if necessary it reminds the person to wear it through the voice interface.

If assisted people own a smartphone, it can be connected to the wearable so that they remain in touch with Arianna even when they are away from home. If needed, they can ask for help and can be easily located using the built-in smartphone sensors.

## 5 Conclusions

Arianna is targeted at collaborative people, who may have mild cognitive or physical disabilities, and may thus exhibit a fair degree of autonomy. Using other types of wearables, Arianna might be extended to other user profiles. As an example, a device embedded in garments could be worn by people with more serious cognitive disorders (who probably would not accept a bracelet) and used to prevent them from leaving the home unnoticed. Child-specific applications are also foreseen.

Arianna is protected by an Italian patent, and international extension has been filed.